\documentclass[twocolumn]{revtex4}
\usepackage{amssymb}
\usepackage{makeidx}
\usepackage{amsmath}
\usepackage{graphicx}
\usepackage{dcolumn}
\usepackage{bm}
\usepackage{epstopdf}

\begin{document}

\title{Electrons diffraction scattering on a traveling wave -- "Inelastic
Kapitza-Dirac effect" }
\author{H.K. Avetissian}
\affiliation{Centre of Strong Fields Physics, Yerevan State University, 1 A. Manukian,
Yerevan 0025, Armenia}
\date{\today }

\begin{abstract}
In this paper conceptual points regarding electrons elastic (Kapitza-Dirac
effect) and inelastic diffraction effect on the different type slowed
electromagnetic wave structures/light gratings are considered. From the
unified point of view it is analyzed the main works on this subject for last
four decades in chronological order, pointing out the essential peculiarity
inherent in induced Cherenkov, Compton, and undulator/wiggler processes too.
This review article has also a goal to resolve confusion in scientific
literature connected with the recently appeared paper \cite{vasya} relating
the electrons diffraction on a traveling wave in a dielectric medium.
\end{abstract}

\pacs{}
\maketitle



In recent years, especially after the realization of experiment on strongly
coherent and intense laser beams \cite{Nature}, the interest to
Kapitza--Dirac effect \cite{Kap-Dir} has been increased. This, in
particular, is conditioned by the significant role of observed electrons
diffraction effect on high-intensity laser beams-gratings \cite{Buch}, \cite%
{Nature}, since the diffracted from highly coherent laser gratings electron
beams are coherent with each other, and Kapitza--Dirac effect is a very
convenient, perhaps the best means to realize coherent electron beams (see,
also, the review paper \cite{Bat}). This is important fact that can serve as
a basis for construction of diverse type new tools, e.g., coherent beam
splitters, new type electron interferometers which would operate at rather
low electron energies etc. Furthermore, since 1975 the scope of the
Kapitza--Dirac effect, which is valid for the electrons elastic scattering
on a standing light wave/grating, has been extended \cite{Analog}, \cite%
{Cher. dif.}, \cite{Und. dif} for inelastic diffraction scattering of
electrons on a slowed travelling wave propagating with the phase velocity $%
\mathrm{v}_{ph}<c$ ($c$ is the light speed in vacuum), to consideration of
general aspects of which is devoted the present review article. The
motivation of the latter have two reasons. First -that is the most
important, is connected with the essential peculiarity inherent in mentioned
processes (induced Cherenkov, Compton, and undulator/wiggler) \cite{Analog}, 
\cite{Cher. dif.}, \cite{Und. dif} -existence of a critical field, above
which a nonlinear threshold phenomenon arises radically changing the
dynamics of electron-slowed travelling wave interaction, and consequently,
the situation for inelastic diffraction scattering. Second, electrons
diffraction effect on a travelling wave in a dielectric medium requires a
special consideration because of a confusion in scientific literature
connected with the appearance recently of a paper on this subject \cite%
{vasya}. Therefore, apart from the pure scientific interest, the current
review has a goal to clear up the-state-of-the-art in the considering field
for last four decades in chronological order, to avoid the further confusion
and misunderstandings in scientific literature.

The predicted by Kapitza and Dirac in 1933 phenomenon of electrons
diffraction scattering on a phase lattice of a standing light wave naturally
for that time had an one-photon character because of the absence of strong
and coherent radiation fields. The experimental situation has been reviewed
by H. Schwarz in \cite{Schwarz,Schwarz2} (see, also \cite{Takeda}). With the
appearance of laser sources Kapitza-Dirac diffraction effect achieved
multiphoton character, the probability of which in the field of strong
counterpropagating laser beams has been done by F. Ehlotzky and C. Leubner
in 1974 on the base of the Helmholtz--Kirchhoff diffraction theory \cite%
{Echl-Lub,Echl-Lub2} (see, also \cite{Fedorov} where a theoretical analysis
is made of the scattering of electrons by a strong standing wave with a
slowly varying amplitude). This is an elastic scattering of electrons moving
in perpendicular direction to wavevectors of counterpropagating waves ($%
\mathbf{k}$ and $-\mathbf{k}$) of the same frequencies -to except the
Doppler shifting because of longitudinal component of electrons' velocity,
and a stationary periodic space-configuration (standing wave structure) is
formed providing the phase matching between the electrons and
counterpropagating waves, only at which the coherent scattering -- electrons
diffraction on a standing light-wave takes place.

Then, in 1975 this phenomenon has been developed by V.M. Haroutunian and
H.K. Avetissian for bichromatic counterpropagating waves and electrons
inelastic diffraction scattering on a slowed interference wave has been
stated \cite{Analog}. In this process, due to the induced Compton effect in
the two wave fields an electron absorbs $s$ photons from the one wave (of
frequency $\omega _{1}$) and coherently radiates $s$ photons into the other
wave (of frequency $\omega _{2}$) and vice versa. This is the condition of
coherency in the induced Compton process corresponding to the resonance
between the Doppler-shifted frequencies in the intrinsic frame of reference
of an electron in the bichromatic counterpropagating waves:%
\begin{equation}
\omega _{1}\left( 1-\frac{\mathrm{v}_{x}}{c}\right) =\omega _{2}\left( 1+%
\frac{\mathrm{v}_{x}}{c}\right)  \label{res}
\end{equation}%
($\mathrm{v}_{x}$ is the electron longitudinal velocity with respect to
counterpropagating waves), at which the conservation of the number of
photons in the induced Compton process takes place, in contrast to
spontaneous Compton effect in the strong wave field where after the
multiphoton absorption a single photon is emitted. The result of such
coherent scattering is equivalent to diffraction of electrons on a slowed
interference wave in the form of a plane traveling wave: 
\begin{equation}
E_{1}E_{2}\cos \left[ \left( \omega _{1}-\omega _{2}\right) \left( t-\frac{%
\omega _{1}+\omega _{2}}{\left\vert \omega _{1}-\omega _{2}\right\vert }%
\frac{x}{c}\right) \right]  \label{wave}
\end{equation}%
($E_{1,2}$ are the amplitudes of the waves' electric field strengths). So,
because of nonstationary field Eq. (\ref{wave}) the energy of an electron
does not conserved (in other words, the energies of absorbed and emitted
photons are different) and in contradistinction to Kapitza--Dirac effect on
a standing phase lattice here the scattering process is inelastic. The
energy change $\Delta \mathcal{E}$ after the interaction is:%
\begin{equation*}
\Delta \mathcal{E}=s\hbar \left( \omega _{1}-\omega _{2}\right) ,\ \Delta
p_{x}=s\hbar \left( \omega _{1}+\omega _{2}\right) /c,
\end{equation*}
\begin{equation}
\ \Delta p_{y}=0;\ s=0,\pm 1,\ldots .  \label{a}
\end{equation}%
(the waves, linearly polarized along the axis $y$, are propagating in the
direction $x$ along which electron acquires momentum transfer $\Delta p_{x}$
in the interaction process; coordinate $z$ is cyclic, hence: $\Delta
p_{z}\equiv 0$) which determines the maximums and minimums of diffraction
peaks over the energy in the probability distribution of inelastic
diffraction of electrons. For the latter, in accordance with Eq. (\ref{res}%
), we shall direct the electron velocity $\mathrm{v}$ at the angle $%
\vartheta _{c}$ to the axis $x$ under the condition:%
\begin{equation}
\mathrm{v}\cos \vartheta _{c}=c\frac{\left\vert \omega _{1}-\omega
_{2}\right\vert }{\omega _{1}+\omega _{2}}  \label{dif}
\end{equation}%
of coherency between two waves. At the condition Eq. (\ref{dif}) the
traveling interference wave Eq. (\ref{wave}) in the intrinsic frame of
reference moving with the electron (of velocity equal to phase velocity of
the slowed wave $\mathrm{v}_{ph}=c\left\vert \omega _{1}-\omega
_{2}\right\vert /(\omega _{1}+\omega _{2})<c$), the travelling wave becomes
a standing phase lattice/light grating, and elastic diffraction scattering
-- Kapitza--Dirac effect occurs. It is obvious that in the laboratory frame
of reference this is inelastic diffraction scattering.

However, the described picture of inelastic diffraction on the slowed
traveling interference wave is only the kinematics of the electrons'
coherent scattering in the induced Compton process with bichromatic
counterpropagating waves. What concerns the dynamics of this interaction, as
has been shown in the works \cite{Cher. Reflec}, \cite{Comp. Reflec}, and 
\cite{Und. Reflec}, the induced coherent processes such Cherenkov, Compton,
and undulator/wiggler ones are possessed with basic peculiarity which
radically changes the ordinary interaction dynamics of an electron with the
periodic wave-field, and above a certain value of the wave intensity a
nonlinear threshold phenomenon of electrons "reflection" or capture by a
slowed traveling wave occurs.\ Hence, we must take into account this fact
and only after the establishment of necessary dynamic condition for
synchronous motion of an electron with the slowed wave one can speak about
the inelastic diffraction effect on a traveling wave.

The exact investigation of the dynamics of induced Compton process reveals
the existence of a critical field in this process%
\begin{equation*}
\xi _{cr}(\omega _{1},\omega _{2})\equiv \left( \xi _{1}+\xi _{2}\right)
_{cr}=
\end{equation*}%
\begin{equation}
\frac{\mathcal{E}_{0}}{mc^{2}}\frac{\left\vert \omega _{1}\left( 1-\frac{%
\mathrm{v}}{c}\right) -\omega _{2}\left( 1+\frac{\mathrm{v}}{c}\right)
\right\vert }{2\sqrt{\omega _{1}\omega _{2}}}  \label{cr}
\end{equation}%
($\xi _{1,2}\equiv eE_{1,2}/mc\omega _{1,2}$ are the dimensionless
relativistic invariant parameters of the waves' intensities, where $e$, $m$
are the charge and mass of the electron). Thus, if the intensity of a slowed
interference wave Eq. (\ref{wave}) exceeds this critical value Eq. (\ref{cr}%
): $\xi _{1}+\xi _{2}>\xi _{cr}(\omega _{1},\omega _{2})$, a nonlinear
threshold phenomenon of particle "reflection" or capture by the slowed
traveling wave occurs \cite{Comp. Reflec}, at which the periodic wave-field
in the intrinsic frame of reference of the slowed interference wave becomes
a potential barrier for a particle, instead of a standing phase lattice, and
reflecting from this barrier electron abandons the wave. Therefore, the
considering effect of electrons diffraction on a travelling wave propagating
with the phase velocity $\mathrm{v}_{ph}<c$ can occur only if the total
intensity of the waves is smaller than the critical value of corresponding
induced process; in case of Compton one --Eq. (\ref{cr}): $\xi _{1}+\xi
_{2}<\xi _{cr}(\omega _{1},\omega _{2})$.

It is necessary to note that the expression of Eq. (\ref{cr}) for critical
field of induced Compton process corresponds to a case $\vartheta _{c}=0$
when it has a simple form and exact analytic solution of this issue is
succeeded. For inelastic diffraction effect that takes place at the angle $%
\vartheta _{c}$ satisfying to condition Eq. (\ref{dif}) of the induced
Compton resonance, the formula of critical field has a relatively complex
form and therefore it is not convenient to present here bulk expressions.
The reader interested in more detailed information regarding the general
cases of electrons "reflection" or capture by a slowed traveling wave at the
arbitrary angle $\vartheta $ can find the corresponding expressions for
critical field in the papers \cite{Cher. mon.}, \cite{Comp. mon.}, and \cite%
{Und. mon.}, or more details in the work \cite{Diss.}.

Electrons inelastic diffraction effect directly on a travelling wave has
been developed in 1976 by H.K. Avetissian \cite{Cher. dif.}, proposing to
use a dielectric medium for decreasing of the wave phase velocity -making it
smaller $c$. It is clear that electron-travelling wave interaction at $%
\mathrm{v}_{ph}<c$ in a dielectric medium corresponds to induced Cherenkov
process \cite{Cher. dif.}. Thus, by the same physical consideration
mentioned above, in \cite{Cher. dif.} has been shown that if one direct the
electron velocity $\mathrm{v}$ at the angle $\vartheta _{ch}$ to the wave
propagation direction%
\begin{equation}
\mathrm{v}\cos \vartheta _{ch}=\frac{c}{n}\text{,}  \label{Ch. cond.}
\end{equation}%
where $n$ is the refractive index of a dielectric medium, under this
condition the travelling wave in the intrinsic frame of reference becomes a
standing phase lattice on which diffraction scattering of electrons occurs.
We will not repeat here the further consideration of inelastic diffraction
effect in a dielectric medium, which has been done in many publications: 
\cite{Cher. dif.}, monographs \cite{2006}, \cite{2016}, and book \cite{Sahak}%
. Only let us to state here the restriction by the wave intensity effect for
inelastic diffraction, representing the formula for critical field of
induced Cherenkov process:%
\begin{equation}
\xi _{cr}(\vartheta )=\frac{c}{2\mathrm{v}}\frac{\mathcal{E}}{mc^{2}}\frac{%
\left( 1-n\frac{\mathrm{v}}{c}\cos \vartheta \right) ^{2}}{\left(
n^{2}-1\right) |\sin \vartheta |}\ ;\qquad \vartheta \neq 0  \label{Ch. cr.}
\end{equation}%
($\mathcal{E}$ is the electron initial energy) that confines the wave
intensity for diffraction effect in a dielectric medium because of the
nonlinear threshold phenomenon of particle "reflection" or capture by a
traveling wave in a dielectric medium \cite{Cher. Reflec}. Consequently, for
"Cherenkov diffraction" effect the wave intensity $\xi $ in a dielectric
medium should be less than the critical value Eq. (\ref{Ch. cr.}): $\xi <\xi
_{cr}(\vartheta )$.

The third case of electrons inelastic diffraction scattering on a slowed
traveling wave, developed by the H.K. Avetissian et al. in 1981 \cite{Und.
dif} again for vacuum case, is the propagation of a plane monochromatic wave
through an electric and magnetic undulator/wiggler. In such space -- time
periodic structures, like to the considered above induced Compton process, a
slowed interference wave is formed, on which diffraction scattering of
electrons occurs by the described scheme, at the intensities below the
corresponding critical values of nonlinear threshold phenomenon of electrons
"reflection" or capture by a travelling wave in the electric and magnetic
undulators (see \cite{Und. Reflec}). As far as descriptions of electrons'
inelastic diffraction scattering on a plane monochromatic wave in the
electric and magnetic undulators are coincide in many features, here we will
consider the more important case of magnetic undulator/wiggler, which is
currently the most perspective coherent tool with extremely large length of
coherency, specifically due to which the x-ray free electron laser has been
realized in the wiggler.

At the propagation of a plane monochromatic wave of frequency $\omega $ and
amplitude of electric field strength $E$ (let of linear polarization, with
relativistic invariant parameter $\xi =eE/mc\omega $) in the linear
undulator with the magnetic field 
\begin{equation}
H_{z}(x)=H\cos \frac{2\pi }{l}x,  \label{und. field}
\end{equation}%
which is characterized by relativistic parameter%
\begin{equation}
\xi _{H}=\frac{elH}{2\pi mc^{2}}  \label{und. par.}
\end{equation}%
($l$ is the space period -undulator step), a slowed interference wave 
\begin{equation}
\xi \xi _{H}\cos \omega \left( t-\frac{c}{1+\frac{\lambda }{l}}\frac{x}{c}%
\right)  \label{und. int. wave}
\end{equation}%
is formed propagating with the phase velocity $\mathrm{v}_{ph}=$ $c/(1+\frac{%
\lambda }{l})<c$, where $\lambda =2\pi c/\omega $ is the wavelength of
monochromatic wave (the geometry is the same as in case of
counterpropagating waves). Hence, if we direct the electron velocity $%
\mathrm{v}$ at the angle $\vartheta _{u}$ to the undulator axis (wave
propagation direction):%
\begin{equation}
\mathrm{v}\cos \vartheta _{u}=\frac{c}{1+\frac{\lambda }{l}},
\label{und. coh.}
\end{equation}%
the slowed interference wave Eq. (\ref{und. int. wave}) in the intrinsic
frame of reference will become a standing phase lattice on which inelastic
diffraction scattering of electrons will occur. In accordance with the
aforementioned dynamic statements for coherent processes, the inelastic
diffraction will take place at the intensities of the total field $\xi +\xi
_{H}$ below the corresponding critical intensity of this process: $\xi +\xi
_{H}<\xi _{cr}\left( \lambda /l\right) $ \cite{Und. Reflec} 
\begin{equation}
\xi _{cr}\left( \frac{\lambda }{l}\right) \equiv \left( \xi +\xi _{H}\right)
_{cr}=\frac{\left\vert 1-\left( 1+\frac{\lambda }{l}\right) \frac{\mathrm{v}%
}{c}\right\vert }{\sqrt{\frac{2\lambda }{l}\left( 1+\frac{\lambda }{2l}%
\right) }}\frac{\mathcal{E}}{mc^{2}}.  \label{und. cr. field}
\end{equation}%
Let us to note that Eq. (\ref{und. cr. field}) corresponds again to a case $%
\vartheta _{u}=0$ for exact nonlinear solution, as was justified above for
bichromatic counterpropagating waves.

It is important to note that the mentioned phenomenon of particle
"reflection" in the induced coherent processes \cite{Cher. Reflec}, \cite%
{Comp. Reflec}, and \cite{Und. Reflec} takes place even in the very weak
wave-fields if the electrons initially are close to the resonance state,
i.e. the electrons' initial longitudinal velocity $\mathrm{v}_{x}$ is close
to the phase velocity $\mathrm{v}_{ph}$ of the slowed wave at which the
critical field is very small too and the condition of $\xi <\xi
_{cr}(\vartheta )$ may be violated \cite{UFN}, \cite{Diss.}. So, no matter
how the traveling electromagnetic (EM) wave is weak, the described inelastic
diffraction effect even on such a weak wave may not occur if the condition $%
\xi <\xi _{cr}(\vartheta )$ is not satisfied.

The considered effect of inelastic diffraction on a traveling wave in all
three induced coherent processes has been described in the scope of
relativistic quantum theory with the help of eikonal wave function -- the
solution of the Klein-Gordon equation for description of multiphoton
processes at the electron-wave nonlinear interaction. As far as the method
of solution of quantum equation of motion in eikonal approximation is
general for described coherent processes, here we will represent only the
dynamic results for inelastic diffraction in case of a traveling EM wave in
a dielectric medium in the original form corresponding to the paper \cite%
{Cher. dif.} and monographs \cite{2006}, \cite{2016} or the book \cite{Sahak}
for comparison its with the result of the paper \cite{vasya} of 2015
pretended to priority for proposed effect of electrons diffraction on a
traveling wave in a dielectric medium.

Thus, neglecting the spin interaction in the quadratic form of Dirac
equation for an electron (as is known, at first, the spin interaction of an
electron with a light-field is rather small compared with the charge
interaction and, second, by it's nature is different than considering
effect; regarding spin dynamics in the Kapitza-Dirac effect, see the paper 
\cite{Keitel}), the latter passes to the Klein-Gordon equation for electron
in the field of a plane EM wave in a dielectric medium: 
\begin{equation*}
-\hbar ^{2}\frac{\partial ^{2}\Psi }{\partial t^{2}}=\left\{ -\hbar ^{2}c^{2}%
\mathbf{\bigtriangledown }^{2}+m^{2}c^{4}+\right.
\end{equation*}%
\begin{equation}
\left. 2ie\hbar c\mathbf{A}(\tau )\mathbf{\bigtriangledown }%
+e^{2}A^{2}\left( \tau \right) \right\} \Psi .  \label{Kl.-Gord}
\end{equation}%
($\hbar $ is the Planck constant, $\mathbf{A}(\tau )$ is the vector
potential and $\tau \equiv t-n\frac{x}{c}$ -wave coordinate of a plane EM
wave). Equation (\ref{Kl.-Gord}) is solved in the eikonal approximation by
electron wave function 
\begin{equation}
\Psi \left( \mathbf{r},t\right) =\sqrt{\frac{N_{0}}{2\mathcal{E}}}f(x,t)\exp %
\left[ \frac{i}{\hbar }\left( \mathbf{pr-}\mathcal{E}t\right) \right] ,
\label{3.91}
\end{equation}%
according to which $f(x,t)$ is a slowly varying function with respect to
free--electron wave function (the latter is normalized on $N_{0}$ particles
per unit volume): 
\begin{equation}
\left\vert \frac{\partial f}{\partial t}\right\vert <<\frac{\mathcal{E}}{%
\hbar }\left\vert f\right\vert ;\qquad \left\vert \frac{\partial f}{\partial
x}\right\vert <<\frac{p_{x}}{\hbar }\left\vert f\right\vert .  \label{3.92}
\end{equation}%
Choosing a concrete polarization of the wave (assume a linear one along the
axis $OY$) and taking into account Eq. (\ref{Kl.-Gord}) for $f(x,t)$ we will
have a differential equation of the first order: 
\begin{equation*}
\frac{\partial f}{\partial t}+\mathrm{v}\cos \vartheta \frac{\partial f}{%
\partial x}=\frac{i}{2\hbar \mathcal{E}}\left[ 2ecp\sin \vartheta \cdot
A_{0}(\tau )\cos \omega \tau \right.
\end{equation*}%
\begin{equation}
\left. -e^{2}A_{0}^{2}(\tau )\cos ^{2}\omega \tau \right] f(x,t),
\label{3.93}
\end{equation}%
where $A_{0}(\tau )$ is a slowly varying amplitude of the vector potential
of quasi-monochromatic wave and $\vartheta $ is the angle between the
electron velocity and the wave propagation direction, as it noted above.
Since should be $\xi _{\max }<\xi _{cr}<<1$ (to be below the ionization
threshold of a dielectric medium too), then for actual values of parameters $%
p\sin \vartheta /mc>>\xi _{\max }$ and the last term $\sim A_{0}^{2}$ in Eq.
(\ref{3.93}) will be neglected. Changing to characteristic coordinates $\tau
^{\prime }=t-x/\mathrm{v}\cos \vartheta $ and $\eta ^{\prime }=t$, it will
be obvious that at the fulfillment of the induced Cherenkov condition $%
\mathrm{v}\cos \vartheta _{ch}=c/n$ Eq. (\ref{Ch. cond.}) the traveling wave
in this frame of coordinates becomes a diffraction lattice over the
coordinate $\tau ^{\prime }$ and for the scattered amplitude of the electron
wave function from Eq. (\ref{3.93}) we have 
\begin{equation}
f(\tau ^{\prime })=\exp \left\{ \frac{iecp\sin \vartheta _{ch}}{\hbar 
\mathcal{E}}\cos \omega \tau ^{\prime }\int_{\eta _{1}}^{\eta _{2}}A(\eta
^{\prime })d\eta ^{\prime }\right\} ,  \label{3.94}
\end{equation}%
where $\eta _{1}$ and $\eta _{2}$ are the moments of the electron entrance
into the wave and exit, respectively. If one returns to coordinates $x$ and $%
t$ and expands the exponential (\ref{3.94}) into a series by Bessel
functions for the total wave function (\ref{3.91}) we will have 
\begin{equation*}
\Psi \left( \mathbf{r},t\right) =\sqrt{\frac{N_{0}}{2\mathcal{E}}}\exp
\left( \frac{i}{\hbar }yp\sin \vartheta _{ch}\right) \sum_{s=-\infty
}^{+\infty }i^{s}J_{s}(\alpha )
\end{equation*}%
\begin{equation}
\times \exp \left[ \frac{i}{\hbar }\left( p\cos \vartheta _{ch}-\frac{%
sn\hbar \omega }{c}\right) x-\frac{i}{\hbar }\left( \mathcal{E}-s\hbar
\omega \right) t\right] ,  \label{3.95}
\end{equation}%
where the argument of the Bessel function 
\begin{equation}
\alpha =\frac{e\mathrm{v}\sin \vartheta _{ch}}{\hbar \omega }%
\int_{t_{1}}^{t_{2}}E(\eta ^{\prime })d\eta ^{\prime },  \label{3.96}
\end{equation}%
and $E$ is the amplitude of the wave electric field strength. The wave
function (\ref{3.95}) describes inelastic diffraction scattering of the
electrons on the slowed traveling wave in a dielectriclike medium. The
electrons' energy and momentum after the scattering are%
\begin{equation*}
\mathcal{E}^{^{\prime }}=\mathcal{E}-s\hbar \omega ,\ \ p_{x}^{^{\prime
}}=p\cos \vartheta _{ch}-\frac{sn\hbar \omega }{c},
\end{equation*}
\begin{equation}
\ \ p_{y}=\mathrm{const};\ \ s=0,\pm 1,\ldots .  \label{3.97}
\end{equation}%
The probability of this process 
\begin{equation}
W_{s}=J_{s}^{2}\left[ \frac{ec^{2}p\sin \vartheta _{ch}}{\hbar \omega 
\mathcal{E}}\int_{t_{1}}^{t_{2}}E(\eta ^{\prime })d\eta ^{\prime }\right] .
\label{3.98}
\end{equation}%
The condition of the applied eikonal approximation (\ref{3.92}) with Eq. (%
\ref{3.94}) is equivalent to the conditions $\left\vert p_{x}^{^{\prime }}%
\mathcal{-}p_{x}\right\vert <<$ $p_{x}$ and $\left\vert \mathcal{E}%
^{^{\prime }}\mathcal{-E}\right\vert <<\mathcal{E}$, which with Eq. (\ref%
{3.97}) gives: $\left\vert s\right\vert n\hbar \omega /c<<p$.

In the case of a monochromatic wave from Eq. (\ref{3.98}) we have 
\begin{equation}
W_{s}=J_{s}^{2}\left( \xi \frac{mc^{2}}{\hbar }\frac{cp\sin \vartheta _{ch}}{%
\mathcal{E}}t_{0}\right) ,  \label{3.99}
\end{equation}%
where $t_{0}=t_{2}-t_{1}$ is the duration of the particle motion in the wave.

As is seen from Eq. (\ref{3.99}) for the actual values of the parameters $%
\alpha >>1$, that is, the process is essentially multiphoton. The most
probable number of absorbed/emitted Cherenkov photons is 
\begin{equation}
\overline{s}\simeq \xi \frac{mc^{2}}{\hbar }\frac{\mathrm{v}}{c}\sin
\vartheta _{ch}\cdot t_{0}.  \label{3.100}
\end{equation}%
The energetic width of the main diffraction maximums $\Gamma (\overline{s}%
)\simeq \overline{s}^{1/3}\hbar \omega _{0}$ and since $\overline{s}>>1$
then $\Gamma (\overline{s})<<\left\vert \mathcal{E}^{^{\prime }}-\mathcal{E}%
\right\vert $.

The scattering angles of the $s$-photon Cherenkov diffraction are determined
by Eq. (\ref{3.97}): 
\begin{equation}
\tan \vartheta _{s}=\frac{sn\hbar \omega \sin \vartheta _{ch}}{cp+sn\hbar
\omega \cos \vartheta _{ch}}.  \label{3.101}
\end{equation}%
From Eq. (\ref{3.101}) it follows that at the inelastic diffraction there is
an asymmetry in the angular distribution of the scattered particle: $%
\left\vert \vartheta _{-s}\right\vert >\vartheta _{s}$, i.e., the main
diffraction maximums are situated at different angles with respect to the
direction of particle initial motion. However, in accordance with the
condition $\left\vert s\right\vert n\hbar \omega /c<<p$ of the eikonal
approximation this asymmetry is negligibly small and for the scattering
angles of the main diffraction maximums from Eq. (\ref{3.101}) we have $%
\vartheta _{-s}\simeq -\vartheta _{s}$. Hence, the main diffraction maximums
will be situated at the angles 
\begin{equation}
\vartheta _{\pm \overline{s}}=\pm \overline{s}\frac{n\hbar \omega }{cp}\sin
\vartheta _{ch}  \label{3.102}
\end{equation}%
with respect to the direction of the particle initial motion.

Note that in 1977 the formula Eq. (\ref{3.99}) has been applied by the
author for explanation of the experiment on energetic widening of an
electron beam at the induced Cherenkov process in a gaseous medium \cite%
{1977}, implemented in SLAC by R. Pantell and his group \cite{Pantell} (see,
also the cited next experiment of this group in 1981, made in the same
conditions).

These results, in particular, inelastic Kapitza--Dirac effect and
diffraction on a traveling wave in a dielectric medium, as was mentioned
above, had been received forty years ago and included in the monographs \cite%
{2006} and \cite{2016}, as well as in the book \cite{Sahak}. Nevertheless,
after the four decades, in 2015 by the same title has been published a paper
in \textit{New Journal of Physics, v. 17, 082002, 2015 }by the authors%
\textit{\ }Armen G Hayrapetyan, Karen K Grigoryan, J\"{o}rg B G\"{o}tte and
Rubik G Petrosyan \cite{vasya}, who report in the Abstract on the
possibility of electrons diffraction scattering on a travelling EM wave in a
dielectric medium: "We report on the possibility of diffracting electrons
from light waves travelling inside a dielectric medium. We show that, in the
frame of reference which moves with the group velocity of light, the
travelling wave acts as a stationary diffraction grating from which
electrons can diffract, similar to the conventional Kapitza--Dirac effect"
(citation from the Abstract). Here only the difference with the known
results is in the physical characteristic "group velocity of light" which is
a rough mistake repeated also in the text, even for a monochromatic wave. As
it has been shown above, diffraction effect is thoroughly the result of the
phase relations and is conditioned exceptionally by the phase velocity of
light that must be smaller than $c$. Beside this incorrectness, authors of
this paper \cite{vasya} ignored the existence of critical field in this
process and influence of considered phenomenon of a particle "reflection" or
capture on the diffraction effect, meanwhile one of these authors is also a
co-author of both inelastic diffraction effect \cite{Und. dif} and
"reflection" phenomenon in the undulators \cite{Und. Reflec}. Concerning the
citation of considering papers devoted to diffraction effect on a travelling
wave, these mechanically are included in the list of References of the paper 
\cite{vasya} in irrelevant context by footnote: "see, for example...".

Concerning the method of calculation of the probability of multiphoton
diffraction scattering, it is well known that there are two adequate methods
- quantum mechanical and Helmholtz--Kirchhoff diffraction theory. The second
method has been proposed and applied by F. Ehlotzky and C. Leubner in 1974
just for the calculation of the multiphoton diffraction probability of
"Kapitza--Dirac effect" in strong laser fields \cite{Echl-Lub}. Following to
this calculation method, authors of the paper \cite{vasya} obtain the known
formula for the probability of multiphoton diffraction scattering in a
dielectric medium Eq. (\ref{3.99}) and claim on the possibility of
diffraction effect in a dielectric medium.

Thus, the probability of $s$-photon diffraction in the paper \cite{vasya} is
given by the formula 16: 
\begin{equation}
\mathcal{I}_{s}=\mathcal{I}_{i}J_{s}^{2}\left( \Delta \right) ;\text{ }%
\Delta =\frac{eA_{0}d}{\hbar c}.  \label{vasya}
\end{equation}%
The analogous expression for $s$-photon diffraction in the paper \cite{Cher.
dif.} is given by the formula 8 (which is the obtained above Eq. (\ref{3.99}%
)):%
\begin{equation}
W_{s}=J_{s}^{2}\left( \frac{ec^{2}E_{0}tP\sin \vartheta _{ch}}{\hbar \omega 
\mathcal{E}}\right) =J_{s}^{2}\left( \frac{eE_{0}d}{\hbar \omega }\right) ,
\label{1}
\end{equation}%
where $d\equiv t\mathrm{v}\sin \vartheta _{ch}$ is the electron-wave
interaction length.

Expressing the amplitude of the wave electric field strength $E_{0}$ by the
amplitude $A_{0}$ of the wave vector potential ($E_{0}=A_{0}\omega /c$), Eq.
(\ref{1}) will have a form: 
\begin{equation}
W_{s}=J_{s}^{2}\left( \Delta \right) ;\ \Delta =\frac{eA_{0}d}{\hbar c}.
\label{2}
\end{equation}%
i.e., the formula (\ref{vasya}) is the same as formula (\ref{1}) or (\ref{2}%
) except of undetermined normalization constant $\mathcal{I}_{i}$, which
should be $\mathcal{I}_{i}=1$ in accordance to the total probability norm of
the process, and general formula for the Bessel functions: 
\begin{equation}
\sum_{s=-\infty }^{\infty }J_{s}^{2}\left( \Delta \right) =1.
\end{equation}
($s<0$ -corresponds to photon radiation, $s>0$ -to photon absorption in the
wave field).

Note that considered effect of electrons inelastic diffraction takes place
at the fulfilment of classical resonance condition Eq. (\ref{Ch. cond.}).
The latter is valid if the quantum recoil of an electron because of photons
absorption-radiation processes can be neglected. Mathematically the quantum
recoil is connected with the second order derivatives of the wave function
which are neglected in the eikonal approximation (see Eq. (\ref{3.92})). For
the diffraction effect with sufficiently long interaction time and large
energy-momentum exchange one should take into account quantum recoil. At
that, instead of the classical Cherenkov condition Eq. (\ref{Ch. cond.}) we
will have the quantum Cherenkov condition with the quantum recoil \cite{2006}%
: 
\begin{equation}
1-n\frac{\mathrm{v}}{c}\cos \vartheta =\frac{s\hbar \omega (n^{2}-1)}{2%
\mathcal{E}}.  \label{3.21}
\end{equation}%
The condition (\ref{3.21}) has transparent physical interpretation in the
intrinsic frame of reference of the slowed wave. In this frame, due to the
conservation of particle energy and transverse momentum the real transitions
in this -strongly quantum regime- occur from a $p_{x}^{\prime }$ state to
the $-p_{x}^{\prime }$ one and we reach the Bragg diffraction effect on a
slowed traveling wave in a dielectric medium at the fulfilment of the
condition: 
\begin{equation}
2p_{x}^{\prime }=s\hbar k^{\prime }\qquad (s=\pm 1,\pm 2...).  \label{3.20}
\end{equation}%
The latter expresses the condition of exact resonance between the particle
de Broglie wave and the \textquotedblleft wave motionless
lattice\textquotedblright . In particular, in this case when the above
mentioned particle capture regime by the slowed traveling wave \cite{Cher.
Reflec} takes place, we have the quantum effect of zone structure of
particle states like the particle states in a crystal lattice, and the
diffraction maxima takes place at the condition Eq. (\ref{3.21}). For the
more details acquaintance with the different regimes of electrons
diffraction on a slowed traveling wave in a dielectric medium we refer the
reader to the works \cite{Q1,Q2,Q3,Q4,Q5}.

\end{document}